\documentclass{aa}  
\usepackage{txfonts}
\usepackage{color}
\usepackage{graphicx}
\usepackage{natbib}
\usepackage{graphicx}
\usepackage{amsmath}
\usepackage{url} 
\usepackage[breaklinks=true,pdfmenubar=true,pdftoolbar=true,colorlinks=true,citecolor=blue,linkcolor=blue,urlcolor=blue]{hyperref}
\bibpunct{(}{)}{;}{a}{}{,} 

\begin{document}

   \title{(sub)Millimeter Emission Lines of Molecules in Born-again Stars\thanks{This publication is based on data acquired with the Atacama Pathfinder 
   		Experiment (APEX) and IRAM 30m telescopes. APEX is a collaboration between the Max-Planck-Institut f\"{u}r Radioastronomie, 
   		the European Southern Observatory, and the Onsala Space Observatory. IRAM is supported by INSU/CNRS (France), MPG (Germany) 
   		and IGN (Spain).}}


   \author{D. Tafoya\inst{1}, J.~A. Toal\'a\inst{2}, W.~H.~T. Vlemmings\inst{1}, M.~A. Guerrero\inst{3}, E. De Beck\inst{1}, M. Gonz\'alez\inst{3},
   	      S. Kimeswenger\inst{4,5}, A.~A. Zijlstra\inst{6}, \'A. S\'anchez-Monge\inst{7}, \and S.~P. Trevi\~no-Morales\inst{8,9}}

   \institute{Chalmers University of Technology, Onsala Space Observatory, 439 92 Onsala, Sweden
   	\protect{\newline}\email{daniel.tafoya@chalmers.se}
     	\and
     	Institute of Astronomy and Astrophysics, Academia Sinica (ASIAA), 10617 Taipei, Taiwan, R.O.C
     	\and
     	Instituto de Astrofísica de Andalucía (IAA-CSIC), Glorieta de la Astronomía s/n, E-18008 Granada, Spain
     	\and
     	Instituto de Astronom{\'i}a, Universidad Cat{\'o}lica del Norte, Av. Angamos 0610, Antofagasta, Chile 
     	\and 
     	Institut f{\"u}r Astro- und Teilchenphysik, Universit{\"a}t Innsbruck, Technikerstr. 25/8, 6020 Innsbruck, Austria
     	\and
     	Jodrell Bank Centre for Astrophysics, School of Physics and Astronomy, University of Manchester, Manchester M13 9PL, UK
     	\and 
     	I. Physikalisches Institut, Universität zu Köln, Zülpicher Str. 77, 50937 Cologne, Germany
     	\and
     	Instituto de Ciencia de Materiales de Madrid, Sor Juana Inés de la Cruz 3, 28049 Cantoblanco, Madrid, Spain
     	\and
     	Observatorio Astronómico Nacional, Apdo. 112, 28803 Alcalá de Henares Madrid, Spain
        	}

   \date{Received 2016; accepted , 2016}

 
  \abstract
   {Born-again stars offer us a unique possibility of studying the evolution of the circumstellar envelope of evolved stars in human 
   	timescales. To present, most of the observations of the circumstellar material in these stars have been limited to study the 
   	relatively hot gas and dust. In other evolved stars, the emission from rotational transitions of molecules, such as CO, is commonly
   	used to study the cool component of their circumstellar envelopes. Thus, the detection and study of molecular gas in born-again 
   	stars would be of great importance to understand their composition and chemical evolution. In addition, the molecular emission 
   	would be an invaluable tool to explore the physical conditions, kinematics and formation of asymmetric structures in the circumstellar 
   	envelopes of these evolved stars. However, until now, all attempts to detect molecular emission from the cool material around 
   	born-again stars have failed.}
   {We searched for emission from rotational transitions of molecules in the hydrogen-deficient circumstellar envelopes of born-again 
   	stars to explore the chemical composition, kinematics, and physical parameters of the relatively cool gas.}
   {We carried out observations using the APEX and IRAM~30m telescopes to search for molecular emission 
   	toward four well studied born-again stars, V4334~Sgr, V605~Aql, A30 and A78, that are thought to represent an evolutionary 
   	sequence.}
   {We detected for the first time emission from HCN and H$^{13}$CN molecules toward \object{V4334~Sgr}, and CO 
   	emission in \object{V605~Aql}. No molecular emission was detected above the noise level toward \object{A30} and \object{A78}. The detected 
   	lines exhibit broad linewidths $\gtrsim$150~km~s$^{-1}$, which indicates that the emission comes from gas ejected during the 
   	born-again event, rather than from the old planetary nebula. A first estimate of the H$^{12}$CN/H$^{13}$CN abundance ratio in the 
   	circumstellar environment of V4334~Sgr is $\approx$3, which is similar to the value of the $^{12}$C/$^{13}$C ratio measured from 
   	other observations. We derived a rotational temperature of $T_{\rm rot}$=13$\pm1$~K, and a total column density of 
   	$N_{{\rm HCN}}$=1.6$\pm0.1\times$10$^{16}$~cm$^{-2}$ for V4334~Sgr. This result sets a lower limit on the amount of 
   	hydrogen that was ejected into the wind during the born-again event of this source. For V605~Aql, we obtained a lower limit 
   	 for the integrated line intensities $I_{^{12}\rm C}$/$I_{^{13}\rm C}$>4.}
   {}

   \keywords{stars: evolution 
   -- stars: AGB and post-AGB 
   -- stars: mass-loss 
   -- stars: circumstellar matter 
   -- stars: carbon
   -- (ISM:) planetary nebulae: individual: A58/V605~Aql, V4334~Sgr, A78, A30
   }

\authorrunning{D. Tafoya et al.}

\maketitle
%

\section{Introduction}

%
\begin{table*}[!ht]
	\caption{Parameters of the observations}             
	\label{Tab:1}      
	\centering                          
	\begin{tabular}{ l c c c c c c c}        
		\hline\hline                 
		Source &Equatorial coordinates&Frequency range&Beam size\tablefootmark{a}  & $\eta_{\rm{MB}}$\tablefootmark{b} &$\eta_{\rm{A}}$\tablefootmark{c} &(1/$\Gamma)$\tablefootmark{d}\\
		(Name)&(h$\,\,$m$\,\,$s\hspace{0.5cm}$\circ \,\, \prime \,\, \prime\prime$)&(GHz)&($\prime\prime$) &&&(Jy~K$^{-1}$) \\      
		\hline            
		\multicolumn{6}{c}{APEX}\\
		\hline 
		V4334~Sgr&17 52 32.69 $\,$ $-$17 41 08.0&352.5 - 356.5&17.3 - 17.1&0.73&0.60& 40.7\\
		&&343.8 - 347.8&17.8 - 17.6&0.73&0.60& 40.7\\
		V605 Aql&19 18 20.47 $\,$ \phantom{+}01 46 59.6&343.8 - 347.8&17.8 - 17.6&0.73&0.60&40.7\\
		&&328.5 - 332.5&18.6 - 18.3&0.73&0.60&40.7\\
		\hline                 
		\multicolumn{6}{c}{IRAM 30m}\\
		\hline 
		V4334~Sgr&17 52 32.69 $\,$ $-$17 41 08.0&228.5 - 236.5&10.8 - 10.4&0.59&0.46&\phantom{0}8.5\\
		&&100.5 - 116.5&24.5 - 21.1&0.78& 0.62&\phantom{0}6.4\\
		&&\phantom{0}84.5 - 100.5&29.1- 24.5&0.81& 0.63&\phantom{0}6.2\\
		A30&08 46 53.49 $\,$ \phantom{+}17 52 46.8&228.5 - 236.5&10.8 - 10.4&0.59&0.46&\phantom{0}8.5\\
		&&100.5 - 116.5&24.5 - 21.1&0.78&0.62&\phantom{0}6.4\\
		&&\phantom{0}84.5 - 100.5&29.1- 24.5&0.81& 0.63&\phantom{0}6.2\\
		A78&21 35 29.38 $\,$ \phantom{+}31  41 45.3&228.5 - 236.5&10.8 - 10.4&0.59&0.46&\phantom{0}8.5\\
		&&100.5 - 116.5&24.5 - 21.1&0.78& 0.62&\phantom{0}6.4\\
		&&\phantom{0}84.5 - 100.5&29.1 - 24.5&0.81& 0.63&\phantom{0}6.2\\
		\hline                                   
	\end{tabular}
	\tablefoot{
		 \tablefoottext{a}{The fullwidth at half maximum (FWHM) of the telescope primary beam can 
			be computed approximately as FWHM$_{\rm{APEX}}$=$17\rlap{.}^{\prime\prime}7\times(345 / [\nu_{\rm obs} /\rm{GHz}])$, and 
			FWHM$_{\rm{IRAM}}$=$10\rlap{.}^{\prime\prime}7\times(230 / [\nu_{\rm obs} /\rm{GHz}])$, where $\nu_{\rm obs}$ is the observation 
			frequency.}
		\tablefoottext{b}{Main beam efficiency of the telescope. The main-beam brightness temperature is obtained as 
			$T_{\rm{MB}}$=$T^{\prime}_{\rm{A}}$/$\eta_{\rm{MB}}$.}
		\tablefoottext{c}{Apperture efficiency of the telescope.}
		\tablefoottext{d}{Inverse of the point source sensitivity of the telescope. The relation between antenna temperature and flux density is 
			expressed as $S_{\nu}$=$T^{\prime}_{\rm{A}}$/$\Gamma$.}
	}
\end{table*}

%

The term \emph{born-again star} is commonly used to refer to a post-Asymptotic Giant Branch (post-AGB) star that temporarily re-visits
the AGB as a consequence of experiencing a final helium shell flash \citep{Iben1984}. During this phase, the born-again star 
undergoes a sudden episode of intense mass-loss and results in the formation of a hydrogen-poor (H-poor) thick circumstellar envelope 
(CSE) that enshrouds the star and renders it invisible at optical wavelengths \citep[e.g.][]{Duerbeck2000}. After most of the gas on top 
of the stellar core is blown away by the strong stellar wind, the surface temperature commences to rise and a tenuous fast stellar wind 
takes over. At this point the star begins to re-trace the post-AGB track, while still burning helium on a thin layer around its core. In a similar 
way to the interacting stellar winds scenario for the creation of a planetary nebula (PN, plural PNe), the fast wind sweeps up the recently 
formed H-poor CSE and creates a compressed shell \citep{Kwok1978}. Once the star becomes hot enough 
($T_{\rm eff}$$\gtrsim$30,000~K), the energetic stellar radiation ionizes the shell and a brand new PN forms \citep[sometimes referred to 
as ``born-again PN''; e.g.][]{Guerrero2012}. If the final helium shell flash occurs when the star is already on the white dwarf cooling 
track, which is known as a very late thermal pulse (VLTP), it is possible that the born-again PN becomes visible while the old PN is still 
detectable. This shows up in the images as a bright compact PN nested inside a faint extended PN \citep[e.g.][]{Jacoby1979}. Observations 
of stars that are undergoing the born-again event have shown that the evolution from the born-again AGB star to the formation of a 
born-again PN happens on a surprisingly rapid timescale of tens of years. This implies that these objects offer us the unique possibility 
of studying the transformation of an AGB star into a PN on human timescales.

There is a small group of sources that are thought to have experienced a VLTP in relatively recent epochs and that currently are at different 
stages of their post-final flash evolution \citep{Zijlstra2002}. The most recent born-gain event was observed in the late 1990's toward the 
central star of a planetary nebula (CSPN) located in the constellation of Sagittarius \citep{Nakano1996,Benetti1996,Duerbeck1996}. The 
object was cataloged with the variable star name V4334~Sgr, but it is commonly known as Sakurai's object, since it was discovered by the 
amateur astronomer Y. Sakurai. Another object that is thought to have experienced a born-again event in a very similar way to Sakurai's 
object, but 75 years earlier, is the star cataloged as V605~Aql, which is the central star of the planetary nebula A58 (PN A66 58; PN G037.5-05.1)
\citep{Wolf1920,Seitter1987,Guerrero1996,Clayton1997}. At present, this star exhibits spectral characteristics of a Wolf-Rayet [WC] CSPN with a 
temperature of $\sim$100~kK and its H-poor CSE has been partially photo-ionized \citep{Clayton2006,van-Hoof2007}. Finally, two post-VLTP 
objects that have already developed a born-again 
PN from the hydrogen-deficient material that was ejected during the born-again AGB phase are the central stars of the PNe A30 (PN A66 30; 
PN G208.5+33.2) and A78 (PN A66 78; PN G081.2-14.9) \citep[][]{Guerrero2012,Toala2015}. It has been estimated that these sources experienced 
a born-again event a thousand years ago \citep{Fang2014}.
 
The four objects described above are often considered in the literature to represent different stages of an evolutionary sequence of 
born-again stars. Moreover, a striking common characteristic of all these objects is that the circumstellar material ejected in the born-again 
event seems to be distributed in an equatorial disk-like structure and a bipolar outflow 
\citep{Borkowski1993,Guerrero1996,Chu1997,Hinkle2008,Chesneau2009,Hinkle2014}. This characteristic is particularly interesting because explaining 
the formation of bipolar structures in evolved stars represents one of the major current challenges in stellar evolution 
\citep{Balick2002,De-Marco2014}. Thus, these objects, which are developing bipolar morphologies in real time, are ideal to study the formation 
of asymmetric PNe. Extensive theoretical work has been devoted to explain the evolution of born-again stars 
\citep[e.g.][and references therein]{Herwig2002,Miller-Bertolami2006a,Fang2014,Woodward2015}. 
In addition, these sources have been observed over a broad range of wavelengths from X-rays to radio waves 
\citep[e.g.][and references therein]{Hajduk2005,Toala2015}. However, to date the detection of molecular emission at millimeter and sub-millimeter 
wavelengths, which traces the cooler molecular component of the CSE, has been elusive. In this paper we present observations carried 
out with the 12m Atacama Pathfinder Experiment (APEX) and the 30~m Institute for Radio Astronomy in the Millimeter Range (IRAM~30m) 
telescopes. We report on the first detection of molecular emission lines toward born-gain stars. This result opens a new window to study in 
real time the evolution of the physical and chemical properties of the hitherto unexplored cool gas around born-again AGB stars.

\begin{table*}
\caption{Detected line emission.}             
\label{Tab:2}      
\centering                          
\begin{tabular}{l c c c c c c c c}        
\hline\hline                 
Molecule        & Transition              & Rest frequency   & Line peak($T_{\rm{MB, peak}}$)\tablefootmark{e,f} &Line peak($S_{\nu,\rm{peak}}$)\tablefootmark{g} & v$_{\rm{LSR}}$\tablefootmark{e,h} & $\Delta$v\tablefootmark{e,i}  & $I$\tablefootmark{j} &rms\tablefootmark{g,k} \\
&                               & (GHz)\phantom{..}          &    (mK) &    (mJy) &   (km~s$^{-1}$)   &       (km~s$^{-1}$)   &    (K~km~s$^{-1}$) & (mJy)  \\      
\hline
\multicolumn{8}{c}{V4334~Sgr}\\
\hline 
HCN & $J$=4$\rightarrow$3 & 354.50548   &16.5$\pm$1.8& 490$\pm$52 & 181$\pm$16&298$\pm$37&4.9$\pm0.8$&139\\
HCN & $J$=1$\rightarrow$0 &  \phantom{0}88.63185   & \phantom{0}6.2$\pm$0.4&31$\pm$2 & 128$\pm$14& 405$\pm$34&2.5$\pm0.3$&\phantom{00}7\\
H$^{13}$CN & $J$=4$\rightarrow$3 & 345.33976   &\phantom{0}7.2$\pm$0.8& 215$\pm$24 & 123$\pm$13&236$\pm$30&1.7$\pm0.3$&\phantom{0}57\\ 
\hline            
\multicolumn{8}{c}{V605 Aql}\\
\hline 
CO    & $J$=3$\rightarrow$2 & 345.79599   &4.0$\pm$0.5&119$\pm$16&96$\pm$11&164$\pm$25&0.7$\pm0.1$&\phantom{0}29 \\      
\hline                                   
\end{tabular}
\tablefoot{
\tablefoottext{e}{Nominal value and standard error obtained from fitting a Gaussian model to the emission line.}\tablefoottext{e}{The 
	main-beam brightness temperature was obtained as $T_{\rm{MB}}$=$T^{\prime}_{\rm{A}}$/$\eta_{\rm{MB}}$, where $T^{\prime}_{\rm{A}}$ 
	is the antenna temperature corrected for the atmospheric attenuation.} 
\tablefoottext{g}{The flux density can be obtained as $S_{\nu}$=$T^{\prime}_{\rm{A}}/\Gamma$ (or equivalently, $S_{\nu}$=$\eta_{\rm{MB}}T_{\rm{MB}}/\Gamma$).}
\tablefoottext{h}{Central LSR velocity.}
\tablefoottext{i}{Fullwidth at half maximum.}
\tablefoottext{j}{Velocity-integrated intensity of the line}
\tablefoottext{k}{Channel-to-channel 1-$\sigma$ rms noise of the line-free channels for a velocity resolution of 20~km~s$^{-1}$.}
}
\end{table*}
%
%
%

\section{Observations}

The observations with APEX were carried out in July and November 2013, and March 2016, toward the sources V4334~Sgr and 
V605~Aql. The typical precipitable water vapor level during the observations was $\sim$1 mm and the total on-source integration time 
was about 1.7 and 5.5 hours for V4334~Sgr and V605~Aql, respectively. The heterodyne receiver APEX-2, which covers the frequency 
range from 267 to 378~GHz, was used as a frontend. In the backend we used the XFFTS spectrometer, which consists of two units that provide an 
instantaneous bandwidth of 2.5 GHz and 32,768 spectral channels each. There is a fixed overlap region of 1.0 GHz between the units, 
hence the effective bandwidth was 4.0 GHz centered at the rest frequency of the CO$\,$($J$=3$\rightarrow$2) and 
HCN$\,$($J$=4$\rightarrow$3) transitions. 

The observations with the IRAM~30m telescope were carried out in May and June 2014. The frontend 
was the Eight MIxer Receiver (EMIR), which covers four bands in the frequency range from $\sim$80 to 360~GHz. The Fast Fourier 
Transform Spectrometer (FTS) was used as a backend. The observations were done in the EMIR bands E090 and E230, covering the frequency 
ranges 84--116~GHz and 228.5--240.5~GHz, respectively, with a spectral resolution of  200~kHz. The setup allowed to simultaneously look 
for emission lines of e.g. CO, $^{13}$CO and HCN. The observations with this telescope included scans toward the four born-again 
stars, V4334~Sgr, V605~Aql, A30 and A78. However, due to an error in the pointing of the observations on V605~Aql, we had to exclude 
those data from the present work.

The temperature scale of the data delivered by the observatories was in units of the antenna temperature corrected for the atmospheric 
attenuation, $T^{\prime}_{\rm{A}}$. In order to compare our observations with other results, the antenna temperature was converted to 
main-beam brightness temperature, $T_{\rm{MB}}$, and to flux density, $S_{\nu}$. The main-beam brightness temperature is obtained as 
$T_{\rm{MB}}$=$T^{\prime}_{\rm{A}}$/$\eta_{\rm{MB}}$, where $\eta_{\rm{MB}}$ is the main beam efficiency.  The relation between antenna 
temperature and flux density is expressed as $S_{\nu}$=$T^{\prime}_{\rm{A}}$/$\Gamma$, where $\Gamma$ is the point-source sensitivity 
of the antenna. The sensitivity can be calculated as [$\Gamma$/K~Jy$^{-1}$]=$(\eta_{\rm{A}}\,[D/\rm{m}]^{2})/3520$, where $D$ is the diameter of the 
telescope and $\eta_{\rm{A}}$ is the aperture efficiency \citep{Wilson2013}. A summary of the observation parameters is presented in Table~\ref{Tab:1}.

The software CLASS of GILDAS\footnote{See \url{http://www.iram.fr/IRAMFR/GILDAS} for more information about the GILDAS softwares.} 
was used to average the spectral scans, remove baselines and smooth the spectral resolution. All the spectra were re-sampled to a 
spectral resolution of 20~km~s$^{-1}$. 

  \begin{figure}
  \centering
   \includegraphics[width=\hsize]{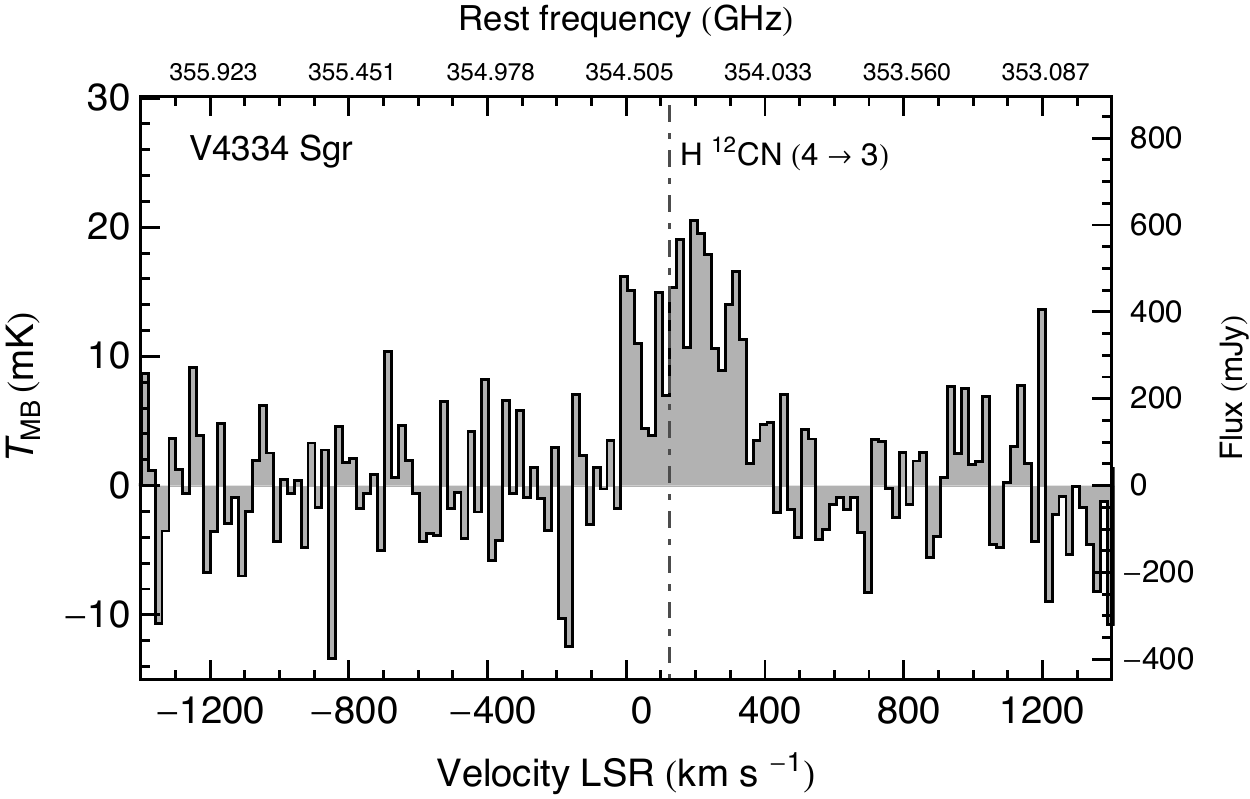}
   \includegraphics[width=\hsize]{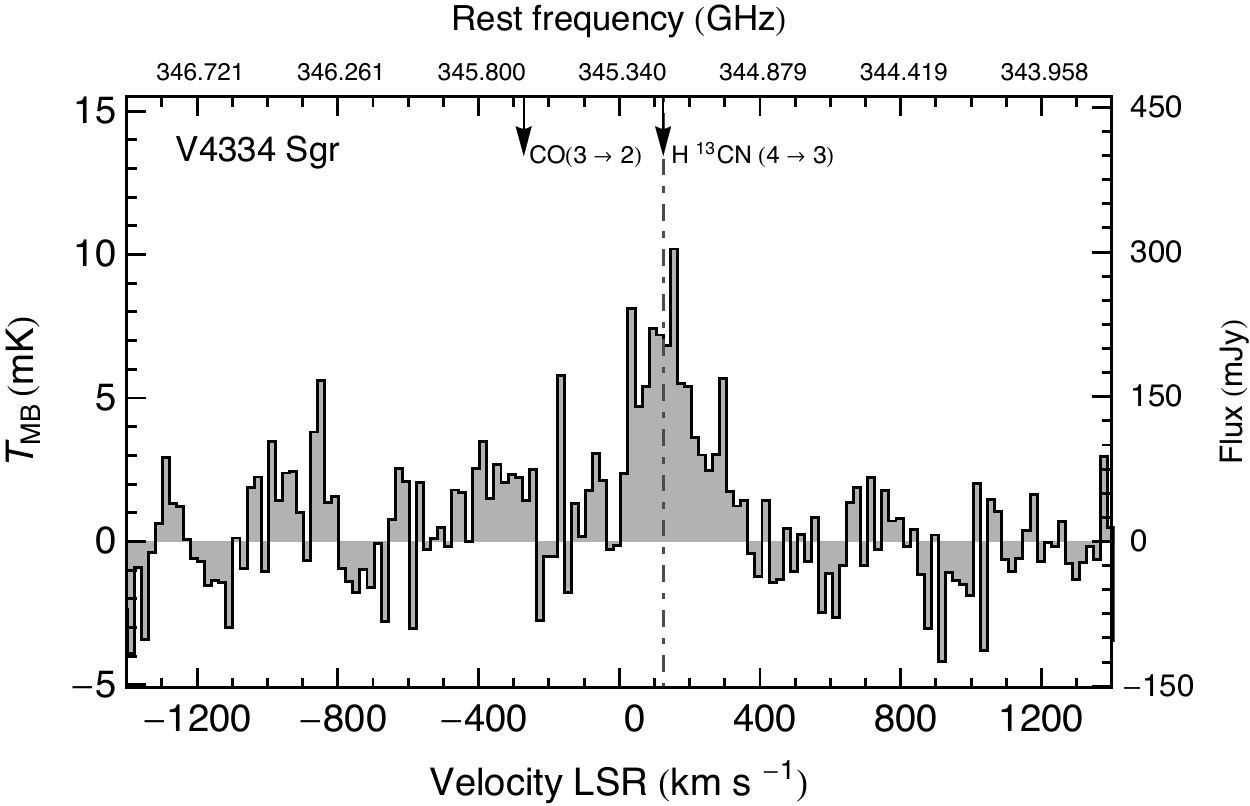}
   \caption{V4334~Sgr spectra of the emission line detected with APEX. 
   The width of the channels is 20~km~s$^{-1}$ and the dotted vertical line corresponds to the systemic velocity of the source, 
   v$_{\rm{sys}}$=125~km~s$^{-1}$ \citep{Duerbeck1997}. The flux density in the right axis was calculated using the corresponding value 
   of $\Gamma$ shown in Table~\ref{Tab:1}. {\bf Top}: Spectrum of the HCN$\,$($J$=4$\rightarrow$3) emission. {\bf Bottom}: 
   Spectrum of the H$^{13}$CN$\,$($J$=4$\rightarrow$3) emission. The arrows in the upper axis indicate the rest frequencies for the 
   CO$\,$($J$=3$\rightarrow$2) and H$^{13}$CN$\,$($J$=4$\rightarrow$3) lines.}
              \label{Fig:1}%
    \end{figure}
   \begin{figure}
  \centering
   \includegraphics[width=\hsize]{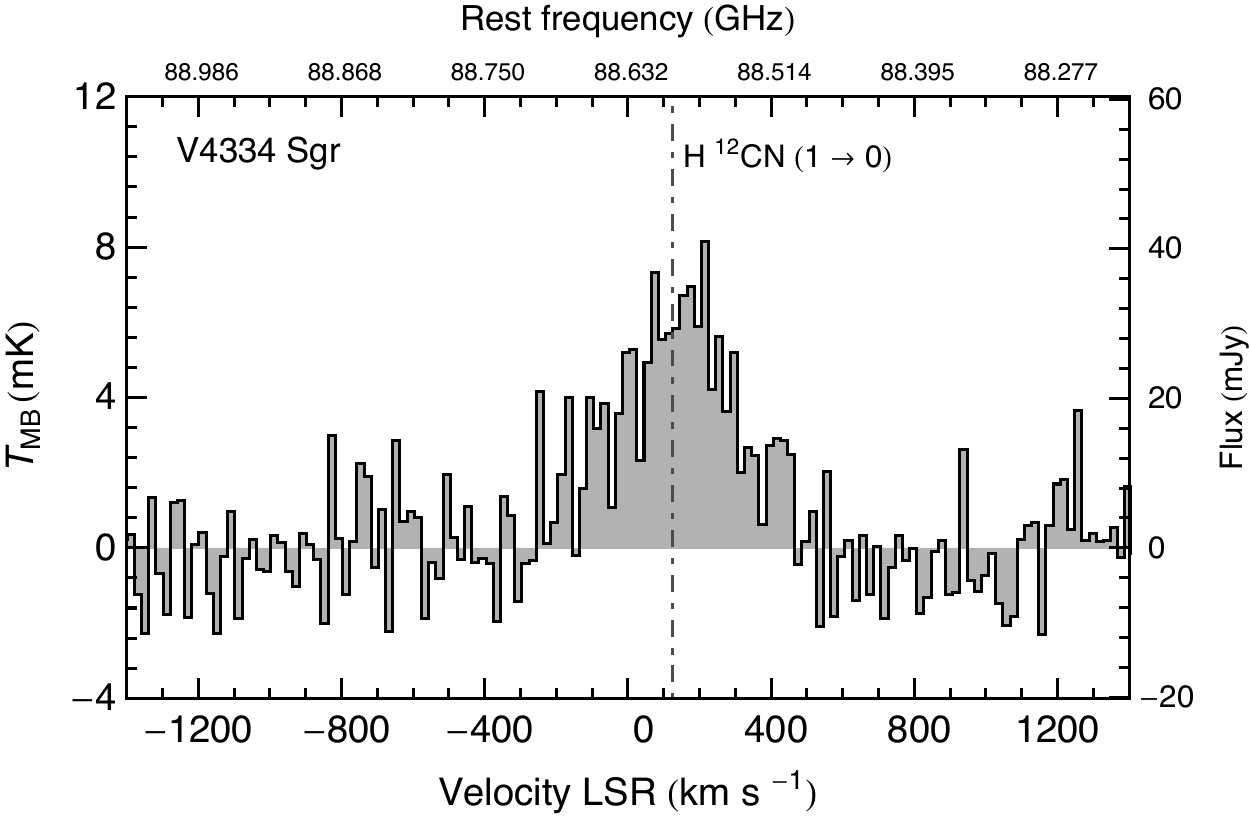}
   \caption{IRAM 30m spectrum of the HCN$\,$($J$=1$\rightarrow$0) emission in V4334~Sgr (Sakurai's object). The width of each
   channel is 20~km~s$^{-1}$. The dotted vertical line corresponds to the systemic velocity of the source, 
   v$_{\rm{sys}}$=125~km~s$^{-1}$ \citep{Duerbeck1997}. The flux density in the right axis was calculated using the corresponding value 
   of $\Gamma$ shown in Table~\ref{Tab:1}.}
              \label{Fig:2}%
    \end{figure}
 \begin{figure}
  \centering
   \includegraphics[width=\hsize]{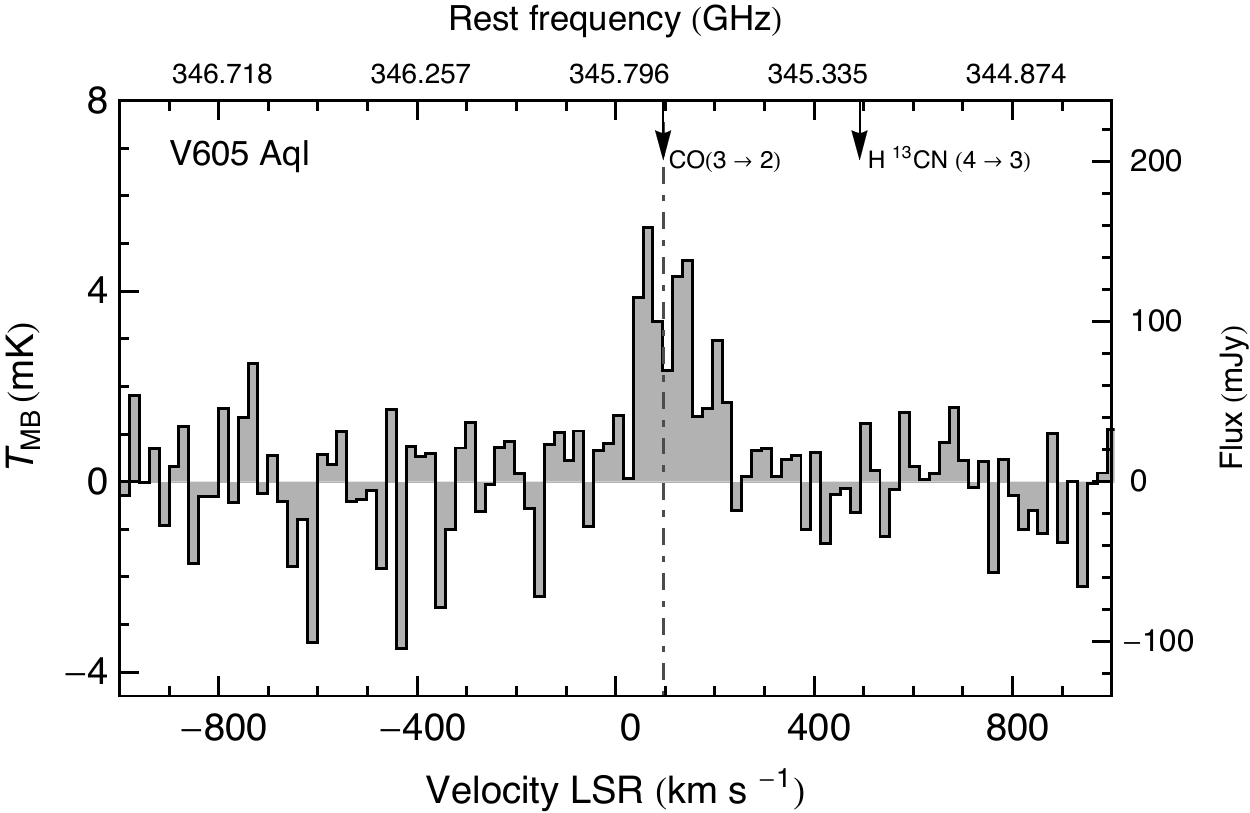}
   \caption{APEX spectrum of the CO$\,$($J$=3$\rightarrow$2) emission in V605~Aql. The width of each channel is 
   	20~km~s$^{-1}$. The dotted vertical line corresponds to the systemic velocity relative to the local standard of rest of the 
   	PN A58, v$_{\rm{sys}}$=96~km~s$^{-1}$, estimated by \citet{Clayton2013}. The arrows indicate the rest frequencies for 
   	the CO$\,$($J$=3$\rightarrow$2) and H$^{13}$CN$\,$($J$=4$\rightarrow$3) lines. The flux density in the right axis was calculated 
   	using the corresponding value of $\Gamma$ shown in Table~\ref{Tab:1}.}
              \label{Fig:3}%
    \end{figure}
%



%

\section{Results}

From our observations we detected, for the first time, molecular line emission toward two of the targeted born-again stars. Emission 
from the molecule HCN and its isotopologue H$^{13}$CN was found in V4334~Sgr. On the other hand, CO$\,$($J$=3$\rightarrow$2) 
emission was detected toward V605~Aql. Although the lines are relatively weak, our deep observations revealed emission of these 
molecules over a broad range of velocities with a signal-to-noise ratio $\gtrsim$3 for the line peaks (see Figs.~\ref{Fig:1}, 
\ref{Fig:2} and \ref{Fig:3}). We fitted Gaussian profiles to the detected lines to obtain their relevant parameters. The best fit values for 
the peak-temperature, $T_{\rm{MB, peak}}$, central velocity, v$_{\rm{LSR}}$, and linewidths, $\Delta$v, are listed in Table~\ref{Tab:2}. 
Table~\ref{Tab:3} shows the channel-to-channel rms noise level of the spectra where no molecular emission was detected. In the 
following we describe the results of the observations for each source.

%
%
\begin{table}
\caption{Noise characteristics for non-detected molecular transitions.}             
\label{Tab:3}      
\centering                          
\begin{tabular}{l c r c}        
\hline\hline                 
Molecule        & Transition              & Rest frequency   & rms\tablefootmark{l}  \\
                    &                               & (GHz) \phantom{..}    &  (mJy)\phantom{..}  \\      
\hline                        
\multicolumn{4}{c}{V4334~Sgr}\\
\hline 
 CO    & $J$=3$\rightarrow$2 & 345.79599   & 57 \\      
 CO    & $J$=2$\rightarrow$1 & 230.53800   & 68\\ 
 CO    & $J$=1$\rightarrow$0 & 115.27120   & 13\\ 
 $^{13}$CO    & $J$=1$\rightarrow$0 & 110.20135   & \phantom{0}6 \\ 
\hline            
\multicolumn{4}{c}{V605~Aql}\\
\hline 
 H$^{13}$CN & $J$=4$\rightarrow$3 & 345.33976   & 29 \\   
 $^{13}$CO & $J$=3$\rightarrow$2 & 330.58797 & 86\\
\hline                                   
\multicolumn{4}{c}{A30}\\
\hline 
CO     & $J$=2$\rightarrow$1 & 230.53800   & 24 \\ 
CO     & $J$=1$\rightarrow$0 & 115.27120   & 16\\ 
$^{13}$CO     & $J$=1$\rightarrow$0 & 110.20135   & \phantom{0}9\\ 
HCN  & $J$=1$\rightarrow$0 &  88.63185   & \phantom{0}7\\ 
\hline                                   
\multicolumn{4}{c}{A78}\\
\hline 
CO     & $J$=1$\rightarrow$0 & 115.27120   & 10\\ 
$^{13}$CO     & $J$=1$\rightarrow$0 & 110.20135   & \phantom{0}6 \\ 
HCN  & $J$=1$\rightarrow$0 &  88.63185  & \phantom{0}6\\ 
\hline                                   
\end{tabular}
\tablefoot{
\tablefoottext{l}{Channel-to-channel 1-$\sigma$ rms noise of the spectrum smoothed to a velocity resolution of 20~km~s$^{-1}$.}
}

\end{table}
%

\subsection{V4334~Sgr (Sakurai's object)}

Among the initial goals of the APEX observations was the detection of CO$\,$($J$=3$\rightarrow$2) emission toward V4334~Sgr, but it was 
not detected above a rms noise level of 57~mJy (Table~\ref{Tab:3}). However, from these observations we serendipitously detected 
emission from the transition $J$=4$\rightarrow$3 of the isotopologue H$^{13}$CN instead, which lies at a frequency $\approx$0.46~GHz 
away from the one of the CO$\,$($J$=3$\rightarrow$2) line (see lower panel of Fig.~\ref{Fig:1} and Table~\ref{Tab:2}). Since the 
H$^{13}$CN$\,$($J$=4$\rightarrow$3) line was detected, we carried out additional observations with APEX to search for 
HCN$\,$($J$=4$\rightarrow$3) emission with the aim of comparing their relative intensities. The HCN$\,$($J$=4$\rightarrow$3) emission 
was successfully detected with a peak-temperature $\approx$2.3 times higher than the one of its isotopologue H$^{13}$CN (upper panel 
of Fig.~\ref{Fig:1}).

From the IRAM~30m telescope observations we also obtained a clear detection of HCN$\,$($J$=1$\rightarrow$0) emission at a frequency 
$\sim$88.6~GHz (Fig.~\ref{Fig:2}). We did not detect CO nor $^{13}$CO emission above the rms noise level from these observations 
(see Table~\ref{Tab:3}).

In general, the central velocities of the HCN and H$^{13}$CN lines, obtained from the Gaussian fits, are in agreement with the 
systemic velocity of the source estimated from other observations, v$_{\rm{sys}}$$=$125~km~s$^{-1}$ \citep{Duerbeck1997}, 
indicating that the emission is indeed associated with the source. It is worth noting that the central velocity of the 
HCN$\,$($J$=4$\rightarrow$3) line differs from the systemic velocity of the source by $\approx$56~km~s$^{-1}$, which 
corresponds to a departure of $\sim$3.5$\sigma$ from the nominal value, given the standard deviation obtained from the Gaussian fit. 
However, since the other lines are consistent with the systemic velocity, it is likely that this discrepancy is due to limitations of the fit 
as a consequence of the lower signal-to-noise ratio of this line.

The widths of the lines show that the emitting material is expanding at a relatively large velocity 
(v$_{\rm exp}$=$\Delta$v/2$\gtrsim$100~km~s$^{-1}$). Similar expansion velocities have been derived from the emission of optical 
lines seen in this source \citep[e.g.][]{Worters2009,Hinkle2014}. We notice that the widths of the HCN$\,$($J$=4$\rightarrow$3) and 
H$^{13}$CN$\,$($J$=4$\rightarrow$3) lines seem to be narrower than the linewidth of the HCN$\,$($J$=1$\rightarrow$0) emission 
(see Figs.~\ref{Fig:1} and \ref{Fig:2} and Table~\ref{Tab:2}). This could be indicating that the region from where the 
HCN$\,$($J$=1$\rightarrow$0) emission is arising, probably cooler gas (i.e. brighter at lower~$J$ transitions), is associated with a 
faster component of the circumstellar envelope. Another possibility is that emission from faint, high-velocity, wings of the 
$J$=4$\rightarrow$3 transitions is not detected due to the higher rms noise of their spectra. 

\subsection{V605~Aql}

Emission from CO$\,$($J$=3$\rightarrow$2) was detected toward V605~Aql using the APEX telescope (see Fig.~\ref{Fig:3} and 
Table~\ref{Tab:2}). The emission is centered at the velocity v$_{\rm LSR}$=96$\pm$11~km~s$^{-1}$. The value is in excellent agreement 
with the systemic velocity derived by \citet{Clayton2013}, v$_{\rm{sys, LSR}}$=96~km~s$^{-1}$, from modeling the broad emission of optical 
lines of [O{\sc iii}] and [N{\sc ii}], attributed to the compact hydrogen-deficient nebula \citep{Pollacco1992}. 

The expansion velocity of the molecular gas, as suggested by the width of the CO line of our observations, is 
v$_{\rm exp}$$\sim$80~km~s$^{-1}$. This is considerably lower than the value of v$_{\rm exp}$$\sim$215~km~s$^{-1}$ derived from 
the spectral modeling of the [O{\sc iii}] and [N{\sc ii}] lines performed by \citet{Clayton2013}. This indicates that the neutral (denser) gas, traced by 
the CO emission, expands at lower velocities than the ionized (less dense) material, as it is actually found in numerical simulations of the 
formation of a born-again PN \citep{Fang2014}.   

No emission of the H$^{13}$CN$\,$($J$=4$\rightarrow$3) line was detected in V605~Aql above the noise level 1--$\sigma$ of 
29~mJy (see Fig.~\ref{Fig:3} and Table~\ref{Tab:3}).

\subsection{A30 and A78}

These sources were observed only using the IRAM~30m telescope. We did not detect any line emission in the frequency bands of our 
observations. The channel-to-channel rms noise level of the spectra of these sources in the frequency ranges corresponding to rotational 
transitions of the molecules CO, $^{13}$CO and HCN is shown in Table~\ref{Tab:3}. 

%
%

\section{Discussion}

\subsection{Molecular emission as a probe of the CSE of born-again stars}

To date most of the studies on the morphology, kinematics and chemical composition of CSEs of born-again-stars are based on 
observations at optical and infrared wavelengths \citep[e.g.][and references therein]{Hinkle2014}. The observations have revealed that 
shortly after the star commences its journey back to the AGB track its chemical composition exhibits dramatic changes \citep[e.g.][]{Eyres1998}. 
This is followed by the condensation of molecular material and dust into a thick circumstellar shell that enshrouds the star, which renders 
it invisible at optical wavelengths. Studies of absorption features in the near and mid-infrared have revealed the presence of molecules such as 
CO, C$_2$, CN, HCN, among others in the CSE of the two youngest born-again stars \citep{Clayton1997,Eyres1998,Pavlenko2004,Evans2006}. 
Infrared emission from dust in globules has also been observed toward A30 and A78 \citep{Borkowski1994,Kimeswenger1998,Phillips2007}, 
but no molecular emission has been reported for these sources.

The emission of the CO molecule has proven to be an invaluable tool to study the morphology and kinematics, as well as the physical 
conditions, of the cool gas around a wide variety of evolved stars, from AGB stars to PNe. Previous searches of CO emission in born-again 
stars at millimeter and sub-millimeter wavelengths yielded inconclusive results and/or non-detections. \citet{Kameswara-Rao1991} reported 
a tentative 2.6-$\sigma$ detection of CO$\,$($J$=1$\rightarrow$0) emission using the Swedish-ESO Submillimetre Telescope (SEST) toward 
V605~Aql. The signal showed a peak main-beam temperature of $T_{\rm MB}$=0.024~K (0.65~Jy) at the position of the LSR velocity of 
138.5~km~s$^{-1}$. These authors concluded that this emission must be originating in the old PN based on the argument that the 
velocity of the hydrogen deficient nebula differs by 60~km~s$^{-1}$ from the LSR of the detected signal. However, given the relatively large expansion 
velocity of the molecular gas ejected during the born-again event, v$_{\rm exp}$$\sim$80~km~s$^{-1}$, as obtained from our observations, 
CO$\,$($J$=1$\rightarrow$0) emission from the hydrogen deficient nebula cannot be ruled out at such LSR velocity. Unfortunately, the spectrum 
of these observations was not published and it is not available in the SEST data archive anymore, preventing us from confirming their 
results. In addition, \citet{Nyman1992} used the SEST and the Onsala 20m Telescope to carry out a survey of CO emission toward a sample of IRAS point 
sources, including V605~Aql. However, their observations toward V605~Aql reached an rms limit of $\sim$1~Jy, which is higher than the peak 
main-beam temperature reported by \citet{Kameswara-Rao1991}, and did not detect any emission. 

On the other hand, soon after V4334~Sgr experienced the VLTP, and just before it disappeared from our view at optical wavelengths, 
\citet{Evans1998} carried out observations with the James Clerk Maxwell Telescope to search for the CO$\,$($J$=2$\rightarrow$1) line toward 
this object, but no emission was detected above an rms level of $\sim$0.65~Jy. To our knowledge, this is the only attempt to search for 
millimeter/submillimeter CO emission toward V4334~Sgr published in the literature. 

Our observations reveal that the molecular emission in the two youngest born-again stars is relatively weak, which explains why previous attempts 
to detect such emission had failed. On the other hand, the non-detection of molecular emission toward the more evolved objects A30 and A78 is in 
agreement with the picture of an evolutionary sequence where the molecules are dissociated as the envelope is photo-ionized by the hot central star.
One particularly interesting result from our observations is the clear detection of HCN emission in V4334~Sgr, while the CO$\,$($J$=3$\rightarrow$2) 
line does not appear in its spectrum (lower panel of Fig.~\ref{Fig:1}). Conversely, CO$\,$($J$=3$\rightarrow$2) 
emission is found in V605~Aql, but no H$^{13}$CN$\,$($J$=4$\rightarrow$3) emission is detected in this source (Fig.~\ref{Fig:3}). It is not clear 
what could be the origin of this difference, leading one to wonder whether this could be due to an evolutionary effect in 
the chemical composition of CSE of these sources, or due to differences between their intrinsic chemical compositions. Particularly, the relatively 
high content of hydrogen-bearing gas in the recently created CSE of V4334~Sgr indicates that a considerable fraction of hydrogen was blown 
out instead of being ingested during the VLTP (see Sect. \ref{column_density}). \citet{Asplund1997} show that the surface hydrogen abundance of 
V4334~Sgr decreased during 1996 and thus the molecules may have formed in the earlier ejecta which still had some hydrogen.

The $^{12}$C/$^{13}$C abundance ratio in the newly formed CSE of V4334~Sgr has been estimated independently by some authors 
\citep{Pavlenko2004,Evans2006,Worters2009}. In general they find a low value of 2$\sim$5, which hints the occurrence 
of a VLTP in this source. From our observations we find that the ratio of the integrated line intensities,
$I_{\rm{H}^{12}\rm{CN}}/I_{\rm{H}^{13}\rm{CN}}$, is $\approx$3 (see Table~\ref{Tab:2}). Under the assumption that both lines arise in the same region 
and that they are optically thin, this ratio translates into the $^{12}$C/$^{13}$C abundance ratio, which would be in agreement with the values 
previously found by other authors. Following a similar argument for V605~Aql, using the upper limit for the $^{13}$CO emission (Table~\ref{Tab:3}), 
we obtain a lower limit for the integrated line intensities $I_{^{12}\rm C}$/$I_{^{13}\rm C}$$>$4. Nonetheless, it is important to remark that the data of the HCN 
and H$^{13}$CN emission toward V4334~Sgr were taken more than 2 years apart and, for sources that evolve so rapidly, one has to be cautious 
when comparing observations that were performed at different epochs.\\

\subsection{Spatial extent of the molecular emission}\label{spatial_extent}

The width of the detected molecular lines, in both sources, show that the emitting material is expanding at relatively large velocities, 
v$_{\rm exp}$$\gtrsim$80~km~s$^{-1}$. This clearly indicates that the observed molecular gas is not associated with the fossil planetary 
nebula, which is expanding at a much lower velocity, v$_{\rm exp}$$\sim$20-30~km~s$^{-1}$, but rather with the material ejected during 
the born-again event. If one assumes that the gas expands at a constant velocity, the angular diameter of the emitting region can be estimated 
as $\theta$=v$_{\rm exp}\,\delta t/D$, where $\delta t$ is the time interval between the born-again event and the observation date, and 
$D$ is the distance to the source. For V605~Aql, the born-again event occurred $\sim$100 years ago and the distance to the source has 
been estimated to be 4.6~kpc \citep{Clayton2013}. Taking half the value of the linewidth of the CO emission as the expansion velocity, 
v$_{\rm exp}$=$\Delta$v$_{\rm CO}$/2, the angular diameter of the emitting region is $\sim$$0\rlap{.}^{\prime\prime}7$. In the case of 
V4334~Sgr the time interval is 20 years; assuming a distance of 3~kpc \citep{Kimeswenger2002}, and calculating the expansion velocity from 
the average of the linewidths of the HCN emission, 
v$_{\rm exp}$=$\left[ \Delta {\rm v}_{{\rm HCN}\,(J=4\rightarrow3)}+\Delta {\rm v}_{{\rm HCN}\,(J=1\rightarrow0)}\right]$/2, the corresponding 
angular diameter is $\sim$$0\rlap{.}^{\prime\prime}5$.

From Hubble Space Telescope (HST) images of the [O{\sc iii}] and [N{\sc ii}] emission, \citet{Wesson2008} measured the diameter of 
the hydrogen deficient nebula in V605~Aql to be $0\rlap{.}^{\prime\prime}76$. In the case of V4334~Sgr, \citet{Hinkle2014} estimate that the 
spatial extent of the material traced by emission of He I at 1.083~$\mu$m might be up to $\sim$$1\rlap{.}^{\prime\prime}4$. However, according 
to the linear relationship for the expansion of the dust shell found by \citet{Evans2006} and \citet{Hinkle2014}, the size of the dust shell should 
be only $\sim$$0\rlap{.}^{\prime\prime}4$ in diameter. From our single dish observations we cannot tell the true distribution of the molecular 
emission, but it can be assumed that the molecular emission comes from a region of similar extent as the structures seen at other wavelengths. 
Interferometric observations with high angular-resolution are necessary to confirm this hypothesis.

\subsection{Column density and rotational temperature of HCN in V4334~Sgr}\label{column_density}

Given that we detected emission of the HCN molecule at two different transitions in V4334~Sgr, we can attempt to estimate the 
rotational temperature and the column density of this molecule in the following way. Assuming LTE conditions, the rotational 
temperature, $T_{\rm rot}$, and total column density, $N$, can be obtained by fitting a linear function in the population diagram 
(also referred to as rotation diagram) to the observational data using the following equation \citep{Goldsmith1999}: 
\begin{equation}\label{Eq:1}
{\rm ln}\frac{N_{u}}{g_{u}}={\rm ln}\frac{N}{Z(T)}-\frac{E_{u}}{kT}, 
\end{equation}
where $N_{u}$ is the column density of molecules in the upper state; $g_{u}$ and $E_{u}$ are the statistical weight and energy of the 
upper level of the transition, respectively; $Z$ is the partition function of the molecule and $k$ is the Boltzmann constant. 

The observed column density of molecules in the upper state, $N_{u}^{\rm obs}$, is obtained as 
\begin{equation}\label{Eq:2}
N_{u}^{\rm obs}=\frac{8\,\pi\,k\,\nu^{2}\,I_{\rm{HCN}}}{h\,c^{3}\,A_{ul}\,f_{\rm s}}\left(\frac{\tau_{\nu}}{1-e^{-\tau_{\nu}}}\right),
\end{equation}
where $\nu$ is the rest frequency of the transition, $h$ is the Planck constant, $c$ is the speed of light in vacuum, $A_{ul}$ is the 
Einstein A-coefficient for spontaneous emission, $I_{\rm{HCN}}$ is the velocity-integrated intensity of the line, which is calculated 
as the product $T_{\rm MB, peak}\times\Delta$v; $f_{\rm s}$ is the beam filling factor and $\tau_{\nu}$ is the optical depth at the 
peak of the line. The latter two parameters are the main source of uncertainty in determining $N_{u}^{\rm obs}$.

In principle, the beam filling factor, defined as the fraction of the area of the beam of the telescope filled by the source, can be estimated 
directly from the angular size of the emitting region, which in the previous subsection was estimated to be 
$\theta_{\rm s}$$\sim$$0\rlap{.}^{\prime\prime}5$ in diameter, and assuming a uniform distribution of the gas. However, a clumpy structure 
of the emitting region would result in a smaller value of the beam filling factor. Furthermore, since the optical depths of the lines are 
not well constrained, we start by assuming that the distribution of the gas is uniform and that the emission of both transitions is 
optically thin, i.e. $\tau_{\nu}$$\ll$1. Later we explore how the results change when these assumptions do not hold.

To calculate $N_{u}^{\rm obs}$ and $E_{u}$ for the two observed lines of the HCN molecule and to perform the linear 
fit using using Equation~\ref{Eq:1}, we used the values of the Einstein A-coefficients, statistical weights and partition functions from the 
Cologne Database for Molecular Spectroscopy \citep{Muller2001,Muller2005}. The fit gives a rotational temperature of 
$T_{\rm rot}$=13$\pm1$~K, and a total column density of $N_{{\rm HCN}}$=1.6$\pm0.1\times$10$^{16}$~cm$^{-2}$. Moreover, 
assuming a distance of 3~kpc and an angular diameter of $0\rlap{.}^{\prime\prime}5$ for this source, we derive a total HCN mass 
$M_{\rm HCN}$=1.4$\pm0.1\times$10$^{-7}$~$M_{\odot}$. \citet{Evans2006} 
analyzed absorption lines of HCN in the infrared and derived values of the temperature and total column density $T_{\rm rot}$=450~K and 
$N_{{\rm HCN}}$=1.4$\times$10$^{17}$~cm$^{-2}$, respectively. The difference between their values and the ones derived from 
our observations is likely due to the fact that the gas probed by the infrared observations lies only in the direction between the infrared 
background and the observer, which is closer to the denser and hotter dust envelope, while we are obtaining an averaged value over the 
whole emitting region, which includes cooler and less dense gas. 

However, it must be pointed out that the values obtained from the analysis above should be considered as a first approximation since they are 
fairly sensitive to the assumptions on the beam filling factor and optical depth of the lines as follows. From Equation~\ref{Eq:1} it can be seen that 
the rotational temperature is inversely proportional to the absolute value of the slope of the line that best fits the data. Accordingly, the total column 
density, $N$, is proportional to the value where it crosses the ordinate axis. On the other hand, from Equation~\ref{Eq:2} it follows that $N_{u}$ is 
inversely proportional to the beam filling factor and linearly proportional to $\tau_{\nu}/(1-e^{-\tau_{\nu}})$. Therefore, different beam filling factors 
and/or optical depths for each transition will cause the values of $N_{u}$ to move upwards and/or downwards in the population diagram, which will 
affect both the slope and the ordinate-crossing point of the best-fitting line. Indeed, as mentioned above, the beam filling factor could be overestimated 
if the emitting region has a rather clumpy distribution. In addition, if the transition of higher energy, HCN$\,$($J$=4$\rightarrow$3), is preferentially 
excited closer to the star the emitting region of this line would be smaller, resulting in a different value of the beam filling factor for each line. 
Furthermore, our assumption of optically thin emission leads to an underestimation of the total column density,  although the agreement between the 
$I_{\rm{H}^{12}\rm{CN}}/I_{\rm{H}^{13}\rm{CN}}$ and the $^{12}$C/$^{13}$C ratios suggests that the lines are indeed optically thin. Hence, it is clear that 
more observations of several transitions at higher angular resolution are necessary to disentangle the physical properties and distribution of the 
molecular gas in this source.

\section{Conclusions}

Using the APEX and IRAM~30m telescopes, we carried out a search for molecular emission toward four sources that are known to have experienced a 
VLTP and detected emission from V4334~Sgr and V605~Aql. This is the first time that millimeter/sub-millimeter molecular emission is detected 
in born-again stars. While the HCN and H$^{13}$CN emission is strong in V4334~Sgr, it does not appear in the spectrum of V605~Aql with the 
same rms noise level. Conversely, CO emission was detected in V605~Aql but not in V4334~Sgr. This result might be pointing to differences in the 
chemical composition of these sources, or to a chemical evolution of their circumstellar envelope. On the other hand, the non-detection of molecular 
emission in the other two, more evolved, born-again stars (A30 and A78) is in agreement with an evolutionary scenario where the molecules are destroyed 
as the central star re-heats and ionizes the CSE. For V4334~Sgr, under the assumptions of LTE and optically thin emission, we derived a rotational 
temperature for the molecular gas of $T_{\rm rot}$=13$\pm1$~K, and a total column density of 
$N_{{\rm HCN}}$=1.6$\pm0.1\times$10$^{16}$~cm$^{-2}$. Considering a distance of 3~kpc and a size of $0\rlap{.}^{\prime\prime}5$, the total 
HCN mass in V4334~Sgr is $M_{\rm HCN}$=1.4$\pm0.1\times$10$^{-7}$~$M_{\odot}$. The ratio of the integrated line intensities, $I_{\rm{H}^{12}\rm{CN}}/I_{\rm{H}^{13}\rm{CN}}$, is $\approx$3 in V4334~Sgr. Assuming that both lines are optically thin, this value represents the $^{12}$C/$^{13}$C abundance ratio, which would be in agreement with previous estimations from infrared observations. Under similar assumptions 
for V605~Aql, we find a lower limit of $I_{^{12}\rm C}$/$I_{^{13}\rm C}$>4. Future observations at  high angular resolution will reveal the distribution 
and kinematics of the gas ejected by these born-again stars. This will be an important step toward understanding the formation of asymmetries in the 
CSE of evolved stars. 
 
\begin{acknowledgements}
The authors thank Karl Torstensson for his valuable help with the observations at the APEX site and to Magnus Persson for helpful discussions. 
The authors also thank the anonymous referee for constructive comments and suggestions that helped improve the manuscript.
The APEX data were obtained via Onsala Space Observatory observing time and are archived with the program identification codes 092.F-9330 and 097.F-9341. DT and WV acknowledge support from ERC consolidator grant 614264. AAZ acknowledges funding from the UK STFC research council. 
\'A.\ S.-M.\ thanks the Deutsche Forschungsgemeinschaft (DFG) for funding support via the collaborative research grant SFB 956, project A6. 
SPTM thanks the Spanish MINECO for funding support from grants AYA2012-32032, CSD2009-00038, FIS2012-32096, and ERC under 
ERC-2013-SyG, G. A. 610256 NANOCOSMOS. 
\end{acknowledgements}

%
%

\bibliography{bibliography_tafoya} 

\end{document}